\documentclass[%
 reprint,
 amsmath,amssymb,
 aps,
]{revtex4-2}

\usepackage{graphicx}
\usepackage{dcolumn}
\usepackage{bm}

\graphicspath{ {./figures/} }
\usepackage{xcolor}

\begin{document}


\title{The application of muon tomography to the imaging of railway tunnels}

\author{L.F.~Thompson}
\email{l.thompson@sheffield.ac.uk}
\author{J.P.~Stowell}
\author{S.J.~Fargher}

\affiliation{Department of Physics and Astronomy, University of Sheffield, Sheffield, S3 7RH, UK.}

\author{C.A.~Steer}
\author{K.L.~Loughney} 
\author{E.M.~O'Sullivan} 

\affiliation{Applied Physics, St Mary’s University, Waldegrave Road, Twickenham TW1 4SX, UK.}

\author{J.G.~Gluyas}
\affiliation{Department of Earth Sciences, Durham University, South Road, Durham, DH1 3LE, UK.}

\author{S.W.~Blaney}
\author{R.J.~Pidcock}
\affiliation{Central Alliance, Alliance House, South Park Way, Wakefield, WF2 0XJ, UK}

\date{\today}

\begin{abstract}

Cosmic Ray Muon Radiography utilizes highly-penetrating cosmic ray muons to image the density profile of an object of interest. 
Here we report on the first trial to use a portable field-deployable cosmic ray tracking system in order to image the whole overburden of a UK railway tunnel with short duration scans (c. 30 minutes). A unknown overburden void was identified and, post-trial, confirmed by railway authorities. These experiments demonstrate the identification of hidden construction shafts with high levels of statistical significance as density anomalies within the data.

\end{abstract}

\maketitle



\section{Railway infrastructure and maintenance}

The United Kingdom's railway network is currently estimated to extend to more than 20,000 miles of track as well as more than 40,000 tunnels, viaducts and bridges. Much of the current infrastructure was constructed in the 19th century and is still in use today \cite{EarlyTunnels}, made possible by a robust safety regime. 

Railway tunnels present significant challenges in assuring their safety over long time periods. Hidden degradation of tunnel linings, where water ingress, differential settlement and other phenomena can lead to so-called voids opening up. 

Similarly, large historic construction shafts, used by the 19th century engineering teams to speed up completion, are also of concern. When a railway tunnel was finished, some vertical construction shafts would be used for ventilation and some sealed over within the tunnel and surface and left. The 1953 hidden shaft collapse in Swinton, near Manchester, England, starkly illustrates that a lack of construction shaft identification and then remediation can cause fatalities for urban populations at the surface \cite{Swinton}. 

Searching for, and monitoring tunnel voids can be time-consuming and may involve many personnel working in potentially hazardous environments. Investigations typically require intrusive drilling into the tunnel lining to discern the presence, or not, of hidden voids and poorly back-filled shafts. Intrusive drilling investigations may cause instabilities and partial failure for marginal linings or, as a shaft is drilled into, the outpouring of collected groundwater. The possibility of an alternate non-invasive technique for rapidly and accurately detecting overburden changes without placing personnel at risk is clearly advantageous. 

Cosmic ray muon radiography poses an ideal solution to this problem, provided imaging systems are capable of meeting the strict timing and portability requirements required for working in live railway tunnels. Cosmic ray muons are also highly penetrating, with significant fluxes capable of passing through hundreds of metres of rock and soil overburden, as a result the technique has previously been successfully applied to the imaging of Eqyptian pyramids \cite{Alvarez,ScanPyramids}, nuclear reactors \cite{morris}, volcanoes \cite{tanaka} and an underground tunnel \cite{guardincerri}.


\section{The Alfreton Old tunnel}
\label{sec:tunnel}

 Situated between Alfreton and Langley Mill in Nottinghamshire, Alfreton Old Tunnel is a disused straight railway tunnel which is 770~m long along its axis from the Langley Mill entrance (53$^{\circ}$05'22.3''N,1$^{\circ}$21'31.5''W) to the Alfreton entrance (53$^{\circ}$05'43.5''N,1$^{\circ}$21'54.6''W). 
 The Old Tunnel was constructed in 1862 and was built using three known construction shafts which were retained for ventilation, these open shafts are visible both from the surface and from inside the tunnel. Alfreton Old Tunnel is sited 28 metres West of the New Tunnel, which was constructed in 1902 and is still in use to the present-day. Unlike the Alfreton New Tunnel, where temporary construction (and subsequently hidden) shafts are well documented~\cite{AlfretonNewTunnel}, no records of temporary construction shafts could be found for the Old Tunnel. 

\begin{figure}
\centering
\includegraphics[width=\linewidth, clip, trim=0 250 0 0]{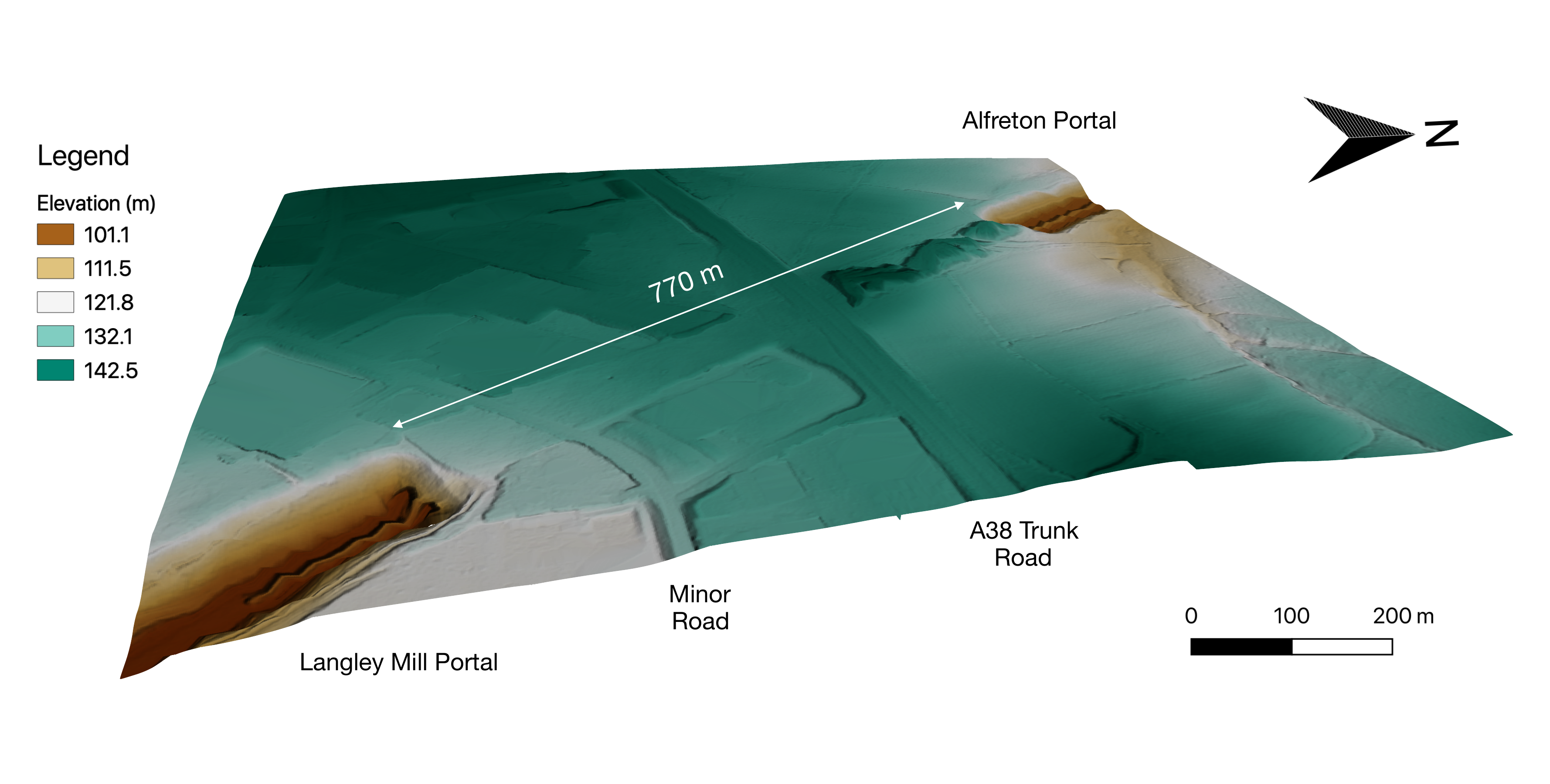}
\caption{Representation of the Alfreton Old and New Tunnels and surrounding using open source LiDAR data \cite{Lidar}.}
\label{fig:TunnelTopology}
\end{figure}

 The area above the tunnel is undeveloped, consisting mainly of scrub growth other than the main A38 dual carriageway and a minor road which both cross over the line of the tunnel. Between the Alfreton entrance and the A38 spoil mounds (from the tunnel's construction) are apparent, up to approximately 7m high. 
Figure~\ref{fig:TunnelTopology} depicts the results of a topographical survey of the tunnels and surrounding area.  Preliminary data from this survey was used in estimates of the tunnel overburden along a straight line directly above the tunnel, which has a maximum, around the middle of the tunnel, of approximately 30~m.

\section{Instrumentation, data acquisition and triggering}
\label{sec:hardware}

The muon tomography system design focused on a number of features key to field deployment including robustness, ease of operation and low power budget. The system comprises 2 horizontal layers of EJ-200 plastic scintillator. The upper layer is segmented into 6 independent rectangular bar detectors 
each 90~cm  $\times$ 15~cm $\times$ 4~cm in size. Similarly the lower layer is segmented into 3 independent square paddle detectors; each 30~cm $\times$ 30~cm $\times$ 4~cm in size. Each individual detector is contained in a light-tight housing and coupled to a photo-multiplier tube for signal readout.  In both layers detectors are placed next to one another with their longest side running perpendicular to the tunnel length and their shortest dimension running vertically. The top layer is fixed 76~cm above the bottom, giving the system a 100$^\circ$ field of view along the axis of the tunnel, and a 76$^\circ$ field of view along its width.

Data from each individual detector is processed using a CAEN DT5740 digitizer,  sampling each channel at 62.5~MHz \cite{CAEN}. The digitizer is triggered from a built-in hardware-based logical OR signal of the lower three detector paddles. The effective detection area of the system is therefore dictated by the area of the combined lower detectors (30~cm $\times$ 90~cm).  
Once a trigger is registered, the signal from each of the 9 detectors is read out and a second software trigger is formed from a logical AND of any upper layer bar with any lower layer paddle within a time window of 64~ns. Specifically, for a detector to be considered in the software trigger it has to observe a signal above a pre-defined threshold. These thresholds are optimised for each detector individually to maximise muon triggering efficiency, whilst rejecting background signals such as dark noise, etc. After threshold optimisation, the estimated efficiency for detecting through-going muons within a single detector was between 88\% and 98\%. For each trigger an event record is saved containing an event time and pulse height for every triggered channel. In effect, this permits 18 different solid angle regions to be sampled simultaneously.

As the system needed to be operated within the tunnel, it was required to be portable, and to be capable of being run remotely with no mains voltage. With a total footprint of 1.3~m $\times$ 1.3~m, the system is small enough to be installed and operated in the back of a commercially available Ford Transit Custom 270 van. Overall the system's power requirements are typically 50~W. Consequently a large capacity battery and DC/AC converter provides sufficient power to run the full system for 50 hours without recharging the battery 

\section{Alfreton Field Trial}

The muon rate drops significantly as a function of overburden, therefore initial feasibility studies were performed to understand the minimum exposure time required to identify voiding or significant density changes with a high reliability. Simulations of muons generated with the CRY library \cite{CRY}, propagated using Geant4 \cite{GEANT} through a 3D reconstruction of the Alfreton Old Tunnel overburden found that exposure times as short as 30 minutes are sufficient to identify open voids and voiding behind the tunnel lining. These simulations were used to inform the required exposure times necessary for scanning the entire length of the Alfreton Old Tunnel.

The Alfreton Old Tunnel data-taking campaign consisted of up to 8 hour shifts for 12 days. Each day the system was transported to the site and assembled in the morning, mimicking operating conditions on a live railway tunnel. Before operation in the tunnel, or in some cases overnight, several open-sky measurements were taken of the unobstructed muon flux through the system for later calibration. During data taking, the system was positioned relative to existing tunnel distance markers that are fixed to the walls at approximately 20~m intervals. A laser range finder was then used to further constrain the system location to within 20~cm of a chosen position along the tunnel length. Similarly, the laser range finder was used to position the system mid-way along the tunnel cross-section to ensure the apex of the tunnel was kept in the centre of the system's field of view. Data were taken along the entire tunnel at 10~m intervals with more than 150 runs being taken in total. For scans around overburden regions of interest the intervals were reduced to 5~m. At each position, 20 or 30 minutes of muon flux data were taken, as well as pressure and temperature readings in 30-second increments. 

\begin{figure*}
\centering
\includegraphics[width=\linewidth, trim=0 40 0 0, clip]{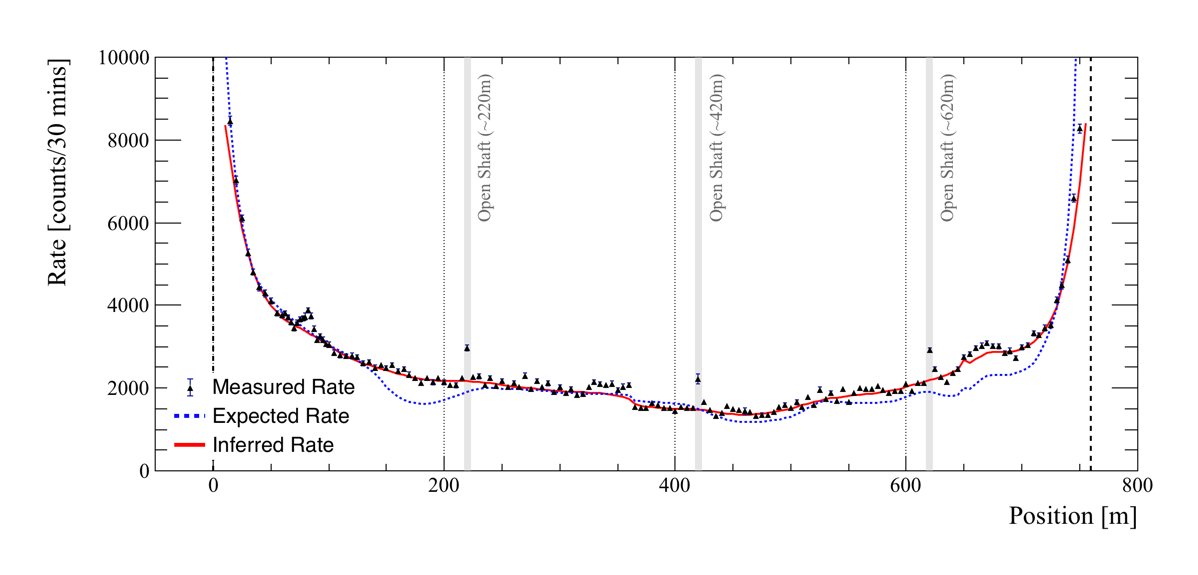}
\caption{Comparison of measured muon rate with the expected rate (from topographical information) and inferred rate along the full length of the tunnel. Distances measured from the Langley Mill entrance.}
\label{fig:muonoverlay}
\end{figure*}

\section{Data analysis and interpretation}

The principal analysis tool comprises simply summing triggers across the full system as described in section \ref{sec:hardware} and converting this to a muon rate per 30 minutes. Figure~\ref{fig:muonoverlay} depicts the variation of this muon rate as a function of distance along the tunnel relative to the Langley Mill portal. The errors displayed on the rate for each data run are purely statistical. The broad features of this figure concur with the expected rate derived from the topographical survey discussed in section \ref{sec:tunnel}. A rapidly reducing muon flux at both tunnel entrances was observed, as the overburden increases, as well as a minimum in the muon flux at 400~m from the Langley Mill entrance where the overburden is the greatest. 

\begin{figure}
\centering
\includegraphics[width=\linewidth, clip, trim=0 0 0 500]{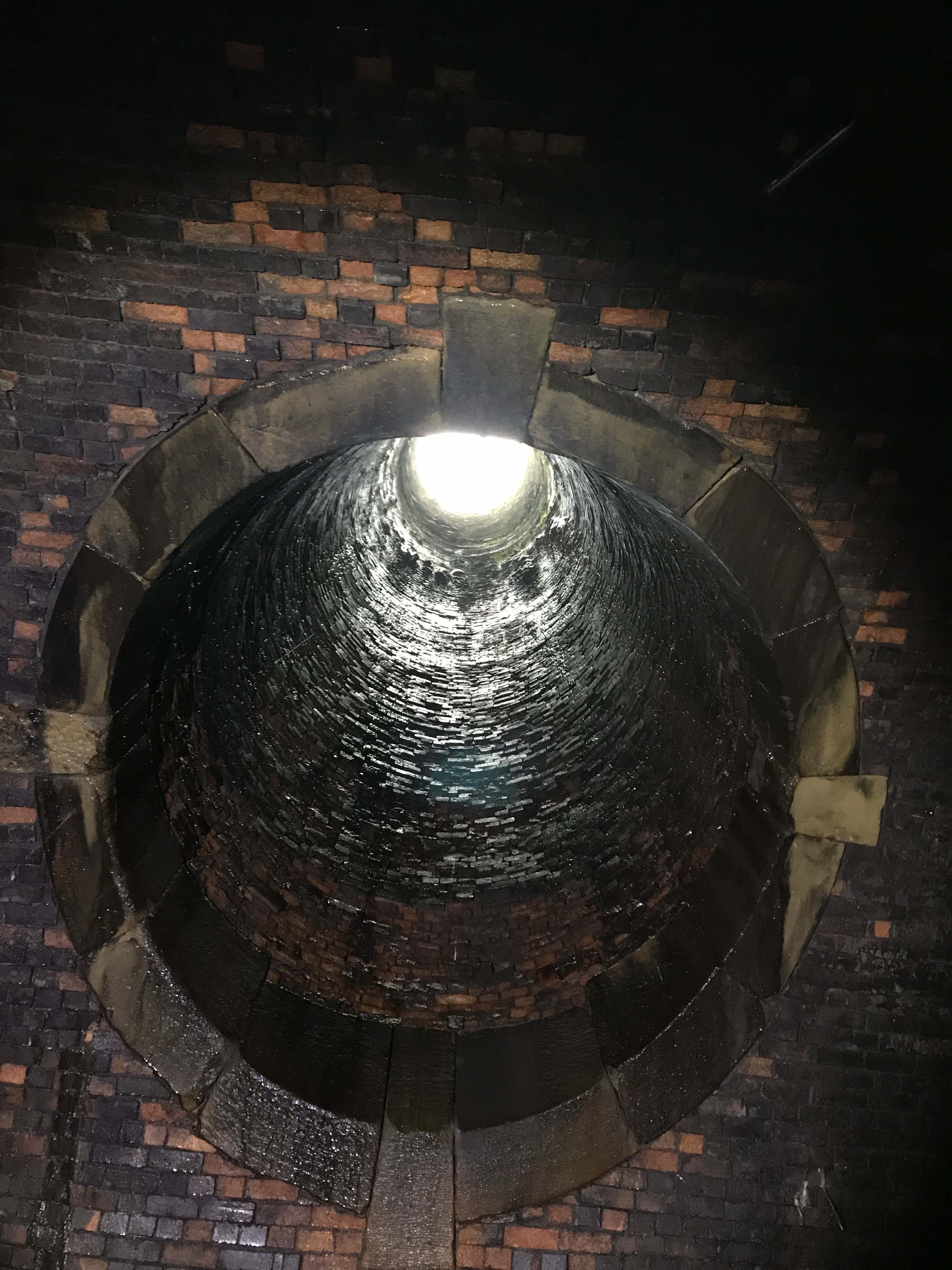}
\caption{Photo of an open ventilation shaft in the Alfreton Old Tunnel, the shaft's diameter is 3~m}
\label{fig:openshaft}
\end{figure}

\begin{figure}
\centering
\includegraphics[width=\linewidth, trim=0 40 0 0, clip]{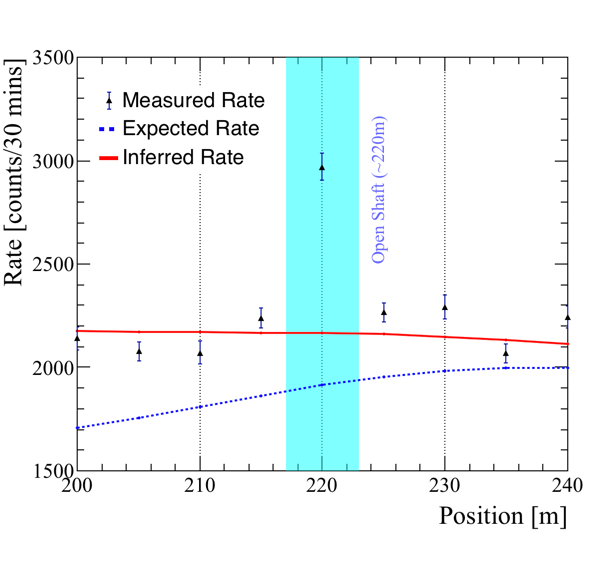}
\caption{Comparison of measured muon rate with the expected rate and inferred rate around the known open shaft at 220~m from the Langley Mill portal end of the tunnel.}
\label{fig:muonzoom220}
\end{figure}

Superimposed on Figure~\ref{fig:muonoverlay} is the known location of the three open shafts (in light grey). Figure~\ref{fig:openshaft} provides a photo of one such open shaft.
The variation in the rate at the open shafts is very clear and highly statistically significant (up to 10 standard deviations when compared to the average trend of the data either side). Any hidden shaft would be expected to create similar variations in the data. Figure~\ref{fig:muonzoom220} provides a zoom of the data around the open shaft at 220~m as an indication of the variation of the flux under the open shafts. Since the shaft is comparable in size to the data taking intervals, a sharp increase is seen when directly under the shaft relative to the points directly either side.

Returning to Figure~\ref{fig:muonoverlay}, whilst there is general agreement, it is clear that there are several areas where there is a disagreement between the expected and measured rate. Note that the discontinuity in the data at 350~m is known to be due to a change in efficiency as a result of exchanging a faulty cable and is not related to an overburden change. Between 150~m and 200~m and also between 600~m and 700~m, the measured muon rate is less than expected from the overburden measurement, indicating a region of higher than expected density/overburden in the ground above the tunnel
After a careful comparison of all possible bar and paddle trigger combinations in this region, it was concluded that the observed discrepancy was due to some larger scale effect and not the presence of a localised high density feature. Since the topography data was sampled in a straight line along the top of the tunnel, it is expected that if additional density variations due to the landscape exist away from this line, but are still within the field of view of the detector, such features would become apparent in the data. This could be corrected for by performing a higher precision topography scan ($\sim$10~m resolution) of the overburden above the tunnel, covering the entire angular field of view of the detector. However, in some cases, where access is limited, or historic knowledge has been lost, such fine scale topography information may be unavailable. Therefore, it was advantageous to consider if muon tomography could be used to identify voiding with only limited access to this information.

The level of redundancy in the system data,  which simultaneously records 18 angular bins, provides the opportunity to address the question of the erroneous expected muon flux from the topographical survey. By making an assumption on the average surrounding rock density, the decrease in muon flux rates compared to the open sky measurement for a given bar and paddle trigger combination can be used to infer the rock thickness along a specific line of sight. Overlaying information from multiple lines of sight and positions can be used to triangulate the average surface height at a given position. The mean surface height for any given position along the tunnel provides an estimate of the overburden given an assumed rock density. In cases where the rock density is not well known, or expected to vary significantly, these inferred overburden estimates can be combined with a limited number of manually measured overburden points to further refine the overburden estimate determined in this way. Furthermore, using a wide binning (e.g. $\sim$20~m) allows the overburden to be estimated with weak dependence on small scale features of interest such as $\sim$5~m shafts. When the overburden is calculated in this way it is possible to infer an updated rate estimate as seen in Figure~\ref{fig:muonoverlay} which removes the large-scale features observed in the previous estimate. This technique is a promising one for general overburden calculations in areas with limited ground-level access.

\begin{figure}
\centering
\includegraphics[width=\linewidth, trim=0 40 0 0, clip]{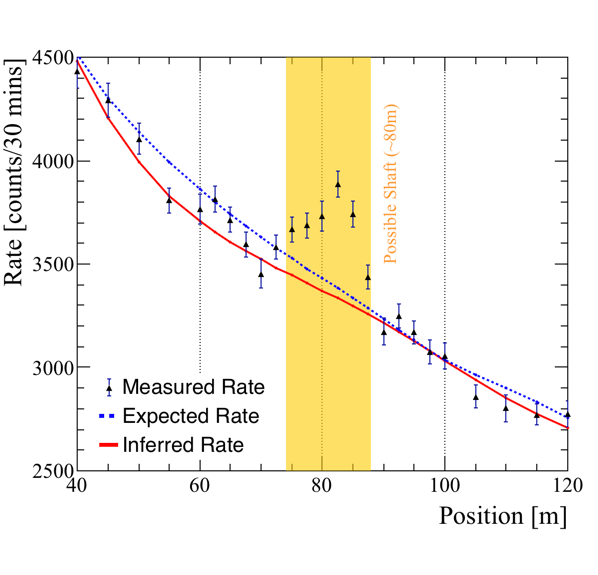}
\caption{Comparison of measured muon rate with the expected rate and inferred rate around the suspected hidden shaft at 80~m}
\label{fig:muonzoom80}
\end{figure}

In addition to the three known open shafts there are additional points along the tunnel where there is poor agreement between observed and inferred muon rates which may be indicative of hidden features that result in overburden changes. One in particular, that at 80~m, was investigated by introducing additional scan points at finer 2.5~m spacings. As illustrated in
Figure~\ref{fig:muonzoom80}, at 80~m there is a statistically significant excess in the muon flux compared with both the expected and inferred flux, however, unlike the case for the open shafts, this feature appears to be broader in extension which may imply a hidden shaft that has undergone some degradation, such as material infall from the sides and/or partial infill. 
Following disclosure of the results of this blind test to the rail authorities, the authors have subsequently been made aware of pre-existing concerns that there is a hidden void in this area.These suspected voided regions are also approximately at the same positions as the known hidden shafts in the Alfreton New Tunnel, providing further confidence. It should be stressed that this information was not made available to the authors at the time of the trial. 

\section{Conclusion}

The first whole-length overburden measurements of a railway tunnel have been performed to search for potentially dangerous voids, such as hidden construction shafts. Cosmic ray muon radiography has been demonstrated as a viable technique for the identification of voiding and significant density changes inside railway tunnel overburden. A portable system with limited angular resolution has been shown to be capable of imaging open shafts with high statistical significance inside the Alfreton Old Tunnel in the UK within a short exposure time of only 100 hours. 
In this blind test, one hidden void above the Alfreton Old Tunnel was identified and, post-trial, confirmed by rail authorities. The use of redundant data to predict the tunnel's overburden illustrates the power of the cosmic ray muon radiography technique as a practical method for overburden mapping even when full topographical information is not available.

\bibliography{alfreton}
\end{document}